\documentclass[lettersize,journal]{IEEEtran}
\usepackage{amsmath,amsfonts}
\usepackage{algorithmic}
\usepackage{algorithm}
\usepackage{array}
\usepackage[caption=false,font=normalsize,labelfont=sf,textfont=sf]{subfig}
\usepackage{textcomp}
\usepackage{stfloats}
\usepackage{booktabs}
\usepackage{url}
\usepackage{verbatim}
\usepackage{graphicx}
\usepackage{cite}
\usepackage{todonotes}
\usepackage{flushend}

\begin{document}

\title{Exploiting and Securing ML Solutions in Near-RT RIC: A Perspective of an xApp}

\author{
    \IEEEauthorblockN{Thusitha Dayaratne\IEEEauthorrefmark{1}, Viet Vo\IEEEauthorrefmark{1} \IEEEauthorrefmark{4}, 
    Shangqi Lai\IEEEauthorrefmark{2}, 
    Sharif Abuadbba\IEEEauthorrefmark{2},  
    Blake Haydon\IEEEauthorrefmark{1}, 
    Hajime Suzuki\IEEEauthorrefmark{2}, 
    Xingliang Yuan\IEEEauthorrefmark{1}\IEEEauthorrefmark{3}, 
    Carsten Rudolph\IEEEauthorrefmark{1}
    }\\
    \IEEEauthorblockA{\IEEEauthorrefmark{1}Monash University, Australia
   }\\
    \IEEEauthorblockA{\IEEEauthorrefmark{2}Data61, CSIRO, Australia
     }\\
     \IEEEauthorblockA{\IEEEauthorrefmark{3}University of Melbourne, Australia
     }
     \\
     \IEEEauthorblockA{\IEEEauthorrefmark{4}Swinburne University of Technology, Australia
     }
}

% The paper headers
\markboth{This is submitted to IEEE Networks}%
{Shell \MakeLowercase{\textit{et al.}}: A Sample Article Using IEEEtran.cls for IEEE Journals}

% \IEEEpubid{0000--0000/00\$00.00~\copyright~2021 IEEE}
% Remember, if you use this you must call \IEEEpubidadjcol in the second
% column for its text to clear the IEEEpubid mark.

\maketitle

\begin{abstract} 
Open Radio Access Networks (O-RAN) are emerging as a disruptive technology, revolutionising traditional mobile network architecture and deployments in the current 5G and the upcoming 6G era. Disaggregation of network architecture, inherent support for AI/ML workflows, cloud-native principles, scalability, and interoperability make O-RAN attractive to network providers for beyond-5G and 6G deployments. Notably, the ability to deploy custom applications, including Machine Learning (ML) solutions as xApps or rApps on the RAN Intelligent Controllers (RICs), has immense potential for network function and resource optimisation. However, the openness, nascent standards, and distributed architecture of O-RAN and RICs introduce numerous vulnerabilities exploitable through multiple attack vectors, which have not yet been fully explored. To address this gap and ensure robust systems before large-scale deployments, this work analyses the security of ML-based applications deployed on the RIC platform. We focus on potential attacks, defence mechanisms, and pave the way for future research towards a more robust RIC platform.

\end{abstract}


\section{Introduction}
Traditional monolithic, proprietary, physical network-based mobile networks are transforming into virtualised functions (virtual network functions and cloud-native functions) that run on commercial off-the-shelf (COTS) servers with fully distributed deployment models with the 5G beyond and emerging 6G standards~\cite{polese2023understanding,liyanage2023open}. This transition enables highly scalable and flexible deployments that can address the variety of use cases independently of any specific vendor. In particular, the introduction of Open Radio Access Networks (O-RAN) and the O-RAN Alliance's standardisation efforts have significantly accelerated this transition in recent years, where there are 14 commercial, 26 field trials, 11 pre-commercial, 14 testing and 12 deployments throughout the globe as of May 2024.

\begin{figure}
    \centering
    \includegraphics[width=\linewidth]{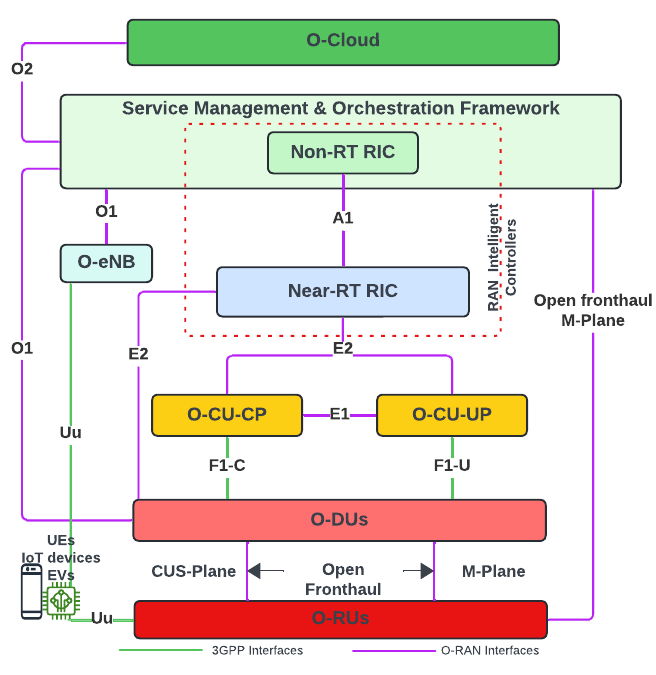}
    \caption{High-level O-RAN Logical Architecture \& Primary Interfaces}
    \label{fig:oran_arch}
\end{figure}

The O-RAN architecture (Figure \ref{fig:oran_arch}) disaggregates the RAN into multiple nodes~\cite{polese2023understanding}, including O-RAN Centralised Units (O-CUs), O-RAN Distributed Units (O-DUs), and O-RAN Radio Units (O-RUs), where each performs a specific task within the network. Existing 3GPP defined protocols and a new set of standardised open interfaces enable the communication among these nodes, allowing network operators to leverage components from different vendors. O-RAN aims to lower deployment costs, improve network performance, and enable more efficient use of resources by leveraging virtualisation, software-defined networking, and cloud technologies. Additionally, O-RAN facilitates the integration of artificial intelligence (AI), machine learning (ML), and network automation into the core RAN functionalities allowing operators to provide enhanced services and experiences to end-users while optimally maintaining their resources. More specifically, O-RAN introduces two Radio Intelligence Controllers (RICs), namely Near-Real-Time RIC (Near-RT RIC) and Non-Real-Time RIC (Non-RT RIC), which provide the native support for AI/ML to further enhance these capabilities. This scalability, flexibility, implicit intelligence, and modularised architecture of O-RAN make O-RAN archirtecture well-suited to address the complexity of upcoming 6G networks with billions of User Equipments (UEs).

Despite the advantages offered by the O-RAN architecture, its inherent openness and distributed architecture introduce a larger attack surface that did not exist with the traditional monolithic architecture~\cite{liyanage2023open}. In particular, multiple vendors and open interfaces create more potential entry points for intruders. Lack of standardisation of security procedures and the nascent nature of RICs can potentially jeopardise the network operations. RICs become a prime target for attackers aiming to disrupt network operations, manipulate traffic, or compromise user privacy, given that RICs interact with multiple network elements and interfaces. Additionally, the integration of AI/ML models into the core network functionalities (through RICs) introduces new vulnerabilities, as malicious actors can exploit these models to manipulate network behaviour or evade detection. Given the emergence of O-RAN as a cornerstone technology for 6G, a thorough understanding of the security posture of these scenarios that involve ML-based xApps is essential before large-scale deployments. In particular, only a limited number of work analysed the security of ML based RIC solutions and potential defence mechanisms~\cite{wisec2024}.

In this work, we analyse the security of ML-based applications (apps) deployed on the Near-RT RIC platform. We specifically focus on outlining the potential attack surface and explore methods to defend these ML-based xApps by using a well-defined O-RAN use case as an example. In particular, we make the following contributions.

\begin{itemize}
    \item Depicts several potential attacks on ML based xApps. Compared to other work that leverages multiple attack vectors, we depict the potential of multitude of attacks from a single attack vector, which makes these attacks more feasible compared to other potential attack vectors (e.g. fake nodes, jammers, compromised devices) that required significant financial investments or significant compromises.
    \item Discuss the potential defence mechanisms to thwart attacks.
    \item Discuss potential future research directions in realising more robust ML solutions against malicious xApps.
\end{itemize}

\section{O-RAN Architecture and RICs}
Figure \ref{fig:oran_arch} depicts the high-level O-RAN architecture and the interfaces among the different O-RAN components. The O-RAN architecture aims to transform traditional proprietary and monolithic RAN architectures into open and disaggregated networks that are empowered by the principles of intelligence and openness. In particular, O-RAN distributes the traditional base station into multiple logical virtualised elements, including O-CU, O-DU, and O-RU, allowing for a more flexible and vendor-neutral network. Further, well-defined set open interfaces and standardised protocols enable multi-vendor interoperability and facilitate the integration of third-party software and hardware components. Furthermore, the O-RAN architecture enables the implementation of network intelligence and automation through the RICs, to cater to diverse service requirements and improve overall network performance. 

RICs provide the intelligence for the O-RAN. RICs leverage open interfaces to communicate and control the O-CUs, O-DUs, and O-RUs. In particular, O-RAN defines two distinct types of RICs based on operational timescales: Near-Real-Time RIC (Near-RT RIC) and Non-Real-Time RIC (Non-RT RIC).
Near-RT RIC operates at the network edge and resides near the O-RUs and O-DU, which allows it to access and process real-time data within the control loop of 10ms - 1s with the support of specialised apps named xApps. Applications that require close to real-time decision-making, including dynamic radio resource management, automated mobility management (cell handovers), and real-time network optimisation leverage the capabilities of the Near-RT RIC and xApps along with the Shared Data Layer (SDL). In contrast, Non-RT RIC operates centrally (as a part of the Service \& Orchestration Management (SMO) platform), capturing the overall status of the network across a predefined geographical area. This placement allows Non-RT RICs to enable network planning and optimisation that execute over the 1s control-loop. 
%More specifically, Non-RT RICs analyse historical data and network trends to make optimal decisions on network-wide congestion control, resource allocation policies, and network slicing policy management with the support of rApps. Additionally, Non-RT RICs can collect and provide enrichment information on O-CUs, O-DUs and O-RUs to Near-RT RICs upon request.

\section{System Model}
We have chosen the traffic steering use case as a generalised example to discuss the security of ML-based xApps in the Near-RT RIC. However, the principles discussed here can be applied broadly to various ML-powered xApps deployed within the O-RAN architecture. These xApps can handle tasks ranging from network anomaly detection to resource optimisation.

\subsection{Use Case: Traffic Steering (TS)}
Mobile network operators treat each UE equally without any customisation, regardless of device capabilities or network capabilities in the current context. In particular, most operators are only executing cell reselection and adjusting the handover parameters. However, given the availability of multiple access technologies (LTE, NR, NR-U, and WiFi), different types of cells (macro, small, and mmWave), types of traffic (data, voice, stream), increasing number of devices, and advanced UE capabilities (support dual connectivity), it is challenging for operators to optimise their network resources to meet customer demands with existing procedures. Hence, it is essential for operators to provide a user-specific customised cellular experience that meets the agreed levels of expectations, while optimally adjusting their load and network conditions. In particular, the O-RAN alliance proposed to leverage a policy based approach to optimally steer the network traffic with the support of ML models \cite{oranusecase}.

The open interfaces and real-time intelligence capabilities of O-RAN enable operators to implement intelligent TS strategies, significantly improving the user experience. The O-RAN alliance recommends a rich set of data \cite{oranusecase} including detailed cell, network utilisation, and user equipment measurements that can be used to achieve this. Successfully TS processes can efficiently distribute the load across the network, maximising resource utilisation and minimising congestion.

\subsection{AI/ML based Traffic Steering}
Given the availability of a rich set of data and the native AI/ML capability of O-RAN architecture allow operators to leverage ML models in achieving optimal traffic management. In particular, several AI/ML-based TS \cite{tamim2023intelligent,kavehmadavani2023intelligent,lacava2023programmable} models were proposed in recent years that employ different AI/ML models, including reinforcement learning, convolutional neural networks, and long-short-term memory networks. Without restricting to a certain AI/ML model, we assume the existence of an AI/ML oracle that leverages SMO, Non-RT RIC, and Near-RT RIC in achieving the desired TS policies. This abstract model aligns with the fourth deployment scenario defined by the O-RAN alliance under AI/ML workflow and requirements, where the combination of Non-RT RIC and Near-RT RIC manage data preparation, training, and model management while Near-RT RIC does the inference. Additionally, the split of the ML model into several rApps and xApps is the approach recommended in the latest O-RAN release. 

\begin{figure}
    \centering
    \includegraphics[width=\linewidth]{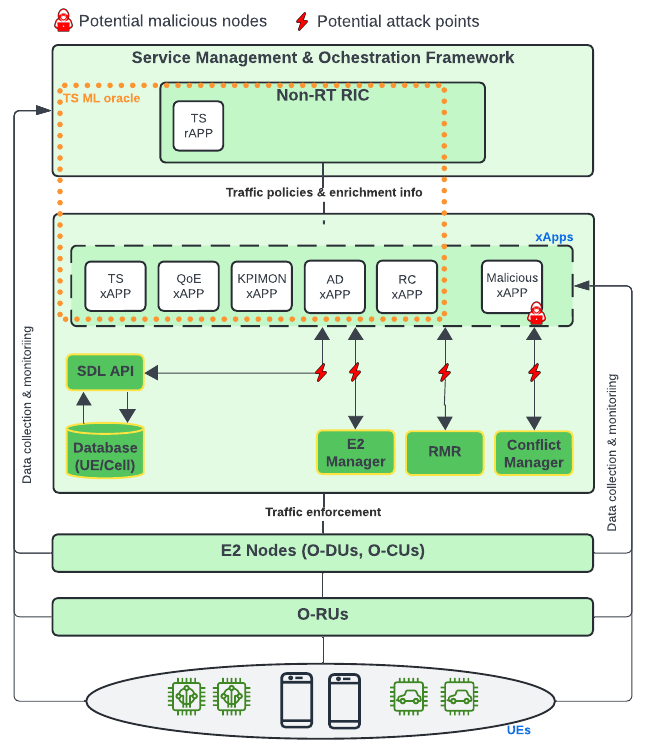}
    \caption{High level system model and threat model}
    \label{fig:system_model}
\end{figure}

Figure \ref{fig:system_model} depicts the abstracted system model. It leverages the functionalities, components, and interfaces of the SMO, Non-RT RIC, and Near-RT RIC with the support of a single rApp and five xApps to collect data, train models, generate policies, enforce policies, and monitor data. 

\textbf{Traffic Steering (TS) rApp:}
\begin{itemize}
    \item The TS rApp is responsible for long-term network monitoring and policy generation for TS.
    \item It gathers various performance metrics at both the UE and cell level through the O1 interface.
    \item Based on this collected data, service level agreements (SLAs), and its ML model, the rApp generates A1 policies. These policies provide guidance on TS preferences for specific UEs to the Near-RT RIC.
\end{itemize}

\textbf{KPIMON (Key Performance Indicator Monitor) xApp:}
\begin{itemize}
    \item The KPIMON xApp gathers radio and system Key Performance Indicator (KPI) metrics from E2 Nodes and stores them in the SDL. This data serves as a valuable resource for other xApps.
\end{itemize}

\textbf{QoE (Quality of Experience) xApp:}
\begin{itemize}
    \item The QoE xApp extracts a relevant set of metrics for a specific UE by querying the UE-Metric and Cell-Metric namespaces within the SDL.
    \item Based on these retrieved metrics, it generates a feature set that captures the UE's current network context.
    \item Utilizing this feature set, the model predicts the UE's future throughput for both the serving cell and any neighboring cells. These throughput predictions are then communicated to the TS xApp to inform potential TS decisions.
\end{itemize}

\textbf{Anomaly Detection (AD) xApp:}
\begin{itemize}
    \item The AD xApp implements a real-time data pipeline that retrieves UE data from the SDL at regular intervals. The retrieved data is then processed to monitor UE metrics for anomalies.
    \item Upon identifying anomalous UE behaviours, the AD xApp triggers alerts and sends the relevant information to the TS xApp.
\end{itemize}

\textbf{RAN Controller (RC) xApp:}
\begin{itemize}
    \item The RC xApp provides the basic implementation of E2 service models compliant with the O-RAN specifications for RN (Radio Network) controllers. It enables sending RIC Control Request messages to RAN/E2 Nodes.
\end{itemize}

\textbf{Traffic Steering (TS) xApp:}
\begin{itemize}
    \item The dedicated TS xApp receives the TS policies generated by the TS rApp.
    \item It leverages the QoE and KPIMON xApps to trigger data collection for specific UEs identified for TS. This involves requesting measurement reports through the E2 interface.
    \item Based on all the gathered information, including model outputs and policies, the xApp issues control actions through the E2 interfaces to steer traffic as needed, including handovers. This is achieved by sending instructions to the RC xApp.
\end{itemize}

\begin{figure}
    \centering
    \includegraphics[width=\linewidth]{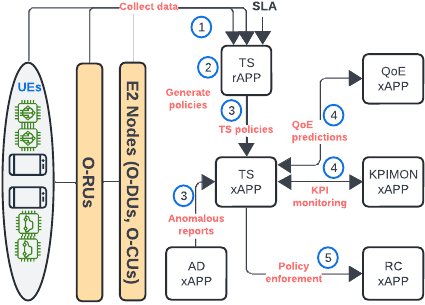}
    \caption{Abstract TS process with the involvement of different apps}
    \label{fig:ts_model}
\end{figure}

The abstracted TS process is shown in Figure \ref{fig:ts_model}. The process begins with acquiring the necessary UE and cell data for the ML model hosted by the TS rApp on the Non-RT RIC. After training the model, the TS rApp generates the required TS policies based on the employed ML model and the SLA. These generated policies are then sent to the TS xApp hosted on the Near-RT RIC, initiating the real-time TS process. Additionally, the AD xApp can also trigger real-time TS upon detecting abnormal traffic, notifying the TS xApp accordingly. Upon activation of the TS xApp, it requests QoE predictions and KPI monitoring outputs from the respective xApps. Leveraging the received policies, reports on anomalous traffic, QoE predictions, and KPI data as inputs to its ML model, the TS xApp generates a specific TS action. Finally, the RC xApp is invoked to execute the derived TS decision.

\section{Threat Model}
O-RAN's disaggregated nature and reliance on RICs introduce an expanded attack surface. This section explores the security vulnerabilities associated with using AI/ML for TS within the O-RAN architecture. In particular, we identify several potential attack vectors that adversaries could exploit to compromise TS mechanisms including fake nodes (O-DU, O-RU), fake UEs, jammers, and malicious xApps. However, in this work we only focus on a single, yet highly plausible, attack vector: malicious xApps deployed on the Near-RT RIC (as depicted in Figure \ref{fig:system_model}). Compared to more resource-intensive attacks (e.g., enroling fake E2 nodes, jamming), deploying malicious xApps requires minimal financial investment, making it a significant threat.

Attackers can leverage the semi-standardised and evolving nature of the xApp platform to deploy malicious xApps on the Near-RT RIC. Security vulnerabilities in the platform or a lack of rigorous evaluation procedures could allow such deployments, similar to how malicious apps infiltrate well-established platforms like iOS and Android app stores despite security measures. We further assume that the malicious xApp has read/write access to the SDL and can subscribe to E2 functionalities, enabling it to issue control signals to manage E2 nodes. These assumptions are well aligned with some of the recent work~\cite{wisec2024,hung2024security} and findings of the O-RAN Working Group 11 study on the security of Near-RT RIC and xApps, where they have mentioned the lack of authentication or authorisation for the databases.

\section{Attacks on ML based Traffic Steering}
Despite the benefits provided by AI/ML-based TS on the O-RAN architecture, this approach has several vulnerabilities under the assumed threat model. We categorise these attacks into adversarial ML attacks and attacks via Near-RT RIC platform based on the specific methods used to execute them.

\subsection{Adversarial ML Attacks}
\subsubsection{Membership Inference Attacks (MIAs)} 
In MIAs, an adversary attempts to extract information regarding the existence of a given input sample in the model's training set~\cite{shi2022membership}. In the context of TS, this could expose sensitive information, such as whether a certain UE was connected to a certain cellular cell or using a certain application, which compromises the privacy. More specifically, if an attacker can infer a UE's presence in the data used to train the QoE model for throughput prediction, they might gain insights into the user's behaviour or location. We depicts two potential MIAs under the TS context.

\paragraph{Leakages for Inference} Upon identifying the target user, the adversary can leverage the malicious xApp's access to the SDL to search for potential residuals of the training data used by the TS model. This allows the adversary to correlate their existing knowledge with the residuals to determine whether the targeted user data was included in the training set.

\paragraph{Poisoning for Inference} An adversary can strategically manipulate the targeted user's data on the SDL. This manipulation could be conducted via the malicious xApp by either injecting fabricated data or manipulating existing data into the SDL. After the manipulation, the adversary needs to observe the outcome of the TS (e.g., predicted throughput, generated traffic steering policies) to determine whether specific user data is being used in the training. The TS output could significantly differ from the training data distribution due to the perturbation.

\textbf{Implication:} The potential impact of a successful MIA on TS could extend beyond privacy violations. For example, having access to knowledge of which user data has been used for TS could allow adversaries to manipulate TS decisions in a chaotic or strategic manner to disrupt the usage of a targeted ordinary user, a reputable or high-profile user, such as a politician, celebrities, etc, or to damage the operator's reputation.

\subsubsection{Model Extraction Attacks (MEAs)}
MEA is another general adversarial ML attack. This attack allows adversaries to replicate the existing ML model by extracting information, including the structure, parameters, and decision boundaries, of the target ML model~\cite{modelextract}. We depict two potential MEAs under the TS context.

\paragraph{Scraping for Extraction} Similar to the MIA, the adversary can leverage the xApp's access to the SDL to gather training data (if it exists) or user data used for real-time inferences. This access could provide insights into the features used by the model and the relationships between these features and the model's outputs, including throughput prediction and traffic steering policies.

\paragraph{Poisoning for Extraction} The adversary can follow the `Poisoning for Inference' approach and observe how the TS model reacts to the manipulated inputs during the learning process. In particular, they can infer details about the TS ML model's logic by analysing the model's outputs of the perturbed data points.

\textbf{Implication:} A successful MEA allows adversaries to replicate the ML-based TS model, causing the network operator to lose its competitive advantage. Furthermore, the replicated model could be used for malicious purposes, such as generating fake traffic or manipulating network performance. Additionally, it can raise privacy concerns, since implicit user-level predictions can reveal sensitive user data. Moreover, adversaries can leverage the model's knowledge (the output of certain traffic situations) to craft user-specific or network-wide attacks that can impact specific users or the network as a whole. In addition, it allows adversaries to craft attacks that can bypass some existing security mechanisms.

\subsubsection{Other Explicit ML Attacks}
In addition to MIAs and MEAs, the adversary could also execute other adversarial attacks such as data poisoning (poisoning training data exploiting the SDL access), model poisoning (exploiting insecure communication between xApps or RMR as explained in the next section), and model evasion attacks (similar to model poisoning by exploiting insecure communication among xApps or RMR). The impact of these attacks could vary from a biased/sub-optimised TS for a targeted user or an unreliable service for a single user to a potential large scale network outage detrimenting the reputation of the operator. 

\subsection{Attacks via Near-RT RIC Platform}
\subsubsection{Exploiting E2 Manager}
Hung et al.~\cite{hung2024security} recently demonstrated that the E2 manager (this is used to manage each E2 Node that is connected to Near-RT RIC via E2 interfaces) uses HTTP protocol instead of HTTPS. The use of HTTP communication allows for eavesdropping and trivial man-in-the-middle attacks. This vulnerability can be exploited by malicious xApps to disrupt AI/ML-based TS, which relies heavily on E2 communication for data collection, policy generation, and enforcement. Attackers could leverage this weakness in several ways, including shutting down E2 nodes, hindering data collection, and effectively preventing the system from making informed TS decisions. Additionally, attackers could potentially alter or prevent the execution of legitimate TS policies by intercepting and manipulating data transmitted over the E2 interface. This could lead to suboptimal network performance or even malicious TS favoring compromised network elements.

\subsubsection{DoSing RIC Message Router (RMR)}
Attackers can use malicious xApps to disrupt communication between legitimate xApps used for TS. The adversaries can send continuous E2 subscription messages to RMR to prevent E2 subscriptions on authentic xApps~\cite{hung2024security}. The RMR is a component within the O-RAN architecture that facilitates communication between xApps by simplifying message exchange. By overwhelming the RMR with subscription requests, attackers can effectively prevent legitimate xApps from subscribing to the data or services they require for TS. This disrupts communication and data exchange between these xApps, hindering their ability to function effectively.

In the context of TS, this attack can significantly disrupt the system's operation, as it relies on collaboration between multiple xApps. When the communication between these xApps is compromised, the system cannot collect the necessary data, make informed decisions, or enforce TS policies. This can lead to degraded network performance and a suboptimal user experience.

\subsubsection{Redirect Attack}
xApps rely on routing tables to establish communication between different xApps and services. These routing tables specify how data should be directed between different xApps. Similar to overwhelming the RMR with subscription requests, attackers can exploit a lack of sender authentication in routing table dissemination~\cite{tseng2023manipulating} to disrupt AI/ML-based TS. 

The adversary can carefully craft the routing table to compromise the essential communication among the four xApps in the context of TS. In particular, fake routing tables could prevent any communication between the xApps, effectively disabling the TS functionality. Further, malicious routing tables could redirect communication towards compromised or spoofed xApps. These fake xApps might provide incorrect or sub-optimal data/decisions to the TS xApp, leading to degraded network performance. Furthermore, disrupted communication between xApps hinders data collection and policy generation, resulting in suboptimal TS decisions. Moreover redirected communication towards compromised xApps could introduce additional security risks including manipulation of TS for malicious purposes.

\subsubsection{Exploit Conflict Manager}
The conflict manager within the Near-RT RIC plays an important role in coordinating decisions from various xApps due to potential conflicts between multiple applications. An adversary could potentially exploit this mechanism through a malicious xApp to launch an exhaustion attack and disrupt the functionality of the ML model~\cite{liyanage2023open,conflict}.

In the context of TS, the malicious xApp can generate a large volume of conflicting TS requests that clash with the authentic requests produced by the TS xApp. This could be achieved by forging user data or manipulating its own output to create false positives. Another potential strategy is to exploit ambiguities in pre-defined rules or a lack of well-defined conflict resolution policies. These strategies can overload the conflict manager and potentially slow down or sub-optimise the TS process, especially if there are significant resource constraints and delays in reevaluations. Since xApps and other Near-RT RIC components share resources in general, an exhaustion attack can also indirectly create a denial-of-service (DoS) effect on the TS ML model as the conflict manager is forced to continuously re-evaluate decisions. However, it is worth noting that a successful exhaustion attack requires the adversary to have a significant understanding of the conflict resolution process and potentially exploit vulnerabilities within the conflict manager's decision logic.

\section{Defence}
The following section discusses three defense mechanisms and presents a potential research question (RQ) for each.

\subsection{AI/ML based Defence}
ML can be employed for anomaly detection to minimise the impact on the discussed attacks on the ML based TS and other ML use cases in general. This involves training an ML model on historical data about legitimate xApp behaviour within the TS ecosystem. The model can then analyse data streams in real-time, including access patterns to the SDL, interactions with E2 nodes, and model outputs generated by xApps. Deviations from expected behaviour, such as unusual data access patterns, attempts to manipulate E2 nodes, or outputs inconsistent with the TS task, could signal a potential malicious xApp. By raising alerts for these anomalies, the ML-based detection system can help prevent attacks like MIAs, MEAs, and E2 node manipulations. This proactive approach can significantly enhance the security posture of O-RAN's traffic steering system.

Another ML based defence is to analyse the source code of deployed xApps. By training a model on a dataset of known malicious and benign xApp code, the system can identify patterns and characteristics, including the APIs and data schemes that a certain app access, indicative of potential threats. This analysis can be performed during the initial deployment stage to prevent potential malicious apps getting deployed at the early stages or to categorise xApps to different risk levels. Additionally, xApp-specific profiles can be developed and continuously monitored to determine any anomalous access patterns. Network operators can introduce additional security checks or restrict access privileges leveraging the determined risk level of the xApp. 

\textbf{RQ:} How to develop efficient and effective AI/ML based solutions to detect potential security breaches and vulnerabilities by fingerprinting xApps?

\subsection{Access Control}
Strict access control mechanisms are another potential solution to mitigate security vulnerabilities associated with malicious xApps. Specifically, unauthorised xApps can be prevented from accessing training data or user data used for real-time inferences by implementing granular access control to the SDL, which limits the information available for an attacker to launch MIAs. Granular control over xApp interactions with E2 nodes can be enforced to prevent unauthorised xApps from issuing manipulative control signals or disrupting communication with E2 nodes.

\textbf{RQ:} How to develop an effective access control mechanisms for xApps within the O-RAN architecture to ensure the robustness of ML-based xApps while balancing latency and other constraints?

\subsection{Zero-Trust}
The adoption of a zero-trust security approach is another potential solution that can significantly enhance defence against malicious xApps. Zero-trust-based approaches~\cite{ramezanpour2022intelligent} have gained momentum in recent years due to their scalability and flexibility in adapting to various use cases. In particular, NIST has recently started to work with the O-RAN Alliance and ATIS to fully incorporate zero-trust architecture into emerging 6G standards. The core principles of zero trust, including assuming breaches, implementing least privilege, and always verifying~\cite{zerotrust}, are well-aligned with an effective defence strategy.

Enforcing the least privilege concept on xApps allows them to access only the specific data and functionalities required for their designated tasks in the context of TS. This approach restricts the information a malicious xApp can access for membership inference or model extraction. Similarly, continuously verifying xApp identity and legitimacy throughout their lifecycle by enforcing authentication and authorisation checks before granting access to resources, along with ongoing monitoring for suspicious behaviour, can help identify and isolate malicious xApps before they can manipulate E2 nodes or steal the ML model. Additionally, implementing microsegmentation to further restrict communication between xApps by creating isolated zones within the Near-RT RIC ensures that malicious xApps cannot easily access sensitive data or manipulate other xApps involved in TS.

\textbf{RQ:} How to leverage the Zero-Trust and its underlying tenets (least privilege, continuous verification and microsegmentation) in creating a more robust security architecture for xApps management in the Near-RT RIC platform?

\section{Discussion \& Future Directions}
Despite the advantages offered by the disaggregation of monolithic cellular networks with the O-RAN architecture, the expanded attack surface can hinder large-scale implementations. Hence, proper identification of potential threats and corresponding defences is essential to realise more robust O-RAN deployments, in particularly for scenarios involving ML based xApps running on the Near-RT RIC platform. This is particularly critical, given the O-RAN's potential to be a cornerstone in managing the complexity, heterogeneity, and implicit intelligence of the upcoming 6G era.

ML models are heavily dependent on the quality of training data. Malicious actors can attempt to compromise the training data used by xApps, leading to biased or manipulated models. Thus, mechanisms to verify data sources and data provenance is essential to prevent these attacks. Techniques such as blockchain-based solutions~\cite{liang2017provchain} can be explored to establish an immutable record of the origin of the data and any modifications made throughout the data pipeline. Additionally, implementing secure communication protocols between data collection points and the Near-RT RIC can further safeguard data integrity. Nevertheless, it is important to ensure that the overhead stays minimal to adhere to the strict time requirements in the Near-RT RIC. 

The adoption of hardware-based Trusted Platform Modules (TPMs) integrated into the Near-RT RIC platform holds significant potential. TPMs provide a secure environment for storing cryptographic keys and performing secure computations. This can be leveraged to ensure the authenticity and integrity of xApps deployed on the platform. Additionally, hardware-based security features can be utilised to isolate xApps from each other, preventing malicious interactions and data breaches. Moreover, TPMs can also be leveraged to realise data integrity and tamper evident data provenance to minimise adversarial ML attacks. However, given the use of containerised cloud environment, more work is required to realise the full potential of available trusted hardware modules. 

Nevertheless, one of the core requirement to be adopted in the Near-RT RIC is its ability to generate near real-time decisions (within 1s) for network operations including the TS. Hence, implementing robust security measures cannot drastically increase the latency. Thus, the optimal balance between security and performance is essential. Techniques including lightweight encryption algorithms and efficient key management protocols must be employed to minimise the performance overhead associated with traditional security measures. Further, available hardware acceleration features, such as FPGA (Field Programmable Gate Arrays), DSP (Digital Signal Processors), ASIC (Application Specific IC), and SoC (System-on-Chip) can be leveraged in implementing security operations without impacting real-time requirements and provide robust isolation.

\section{Conclusion}
This work analysed the security of ML-based xApps deployed within the O-RAN architecture considering its suitability for managing the anticipated complexity and requirements of beyond 5G and 6G eras.. We leveraged a highly plausible attack vector, a malicious xApp, to highlight the potential attack surface, focusing on a well-defined O-RAN traffic steering use case. Additionally, we explored potential defence mechanisms to mitigate these threats. Finally, we discussed potential future research directions to further strengthen the robustness of ML solutions against malicious xApps, paving the way for secure and reliable deployments within the O-RAN ecosystem.

% \begin{table}[]
% \centering
% \begin{tabular}{p{0.5\linewidth}  p{0.15\linewidth} p{0.1\linewidth}}
% \toprule
% \textbf{Objective}                & \textbf{Entities}            & \textbf{ML Model} \\
% \midrule
% Traffic steering\cite{kavehmadavani2023intelligent,lacava2023programmable}             & SMO, Non-RT RIC, Near-RT RIC & RL, CNN           \\
% Rogue base station detection\cite{huang_developing_2023} & UE, Near-RT RIC              & kNN, SVM, RF      \\
% Inference classification\cite{noauthor_system-level_2024,sapavath_experimental_2023}     & Near-RT RIC                  & DNN               \\
% Slice management\cite{9927252}             & SMO, Non-RT RIC, Near-RT RIC & LSTM, RL   \\
% \bottomrule
% \end{tabular}
% \caption{Recent work on the use of ML in RICs}
% \end{table}
\section*{Acknowledgement}
This research paper is conducted under the 6G Security Research and Development Project, as led by the Commonwealth Scientific and Industrial Research Organisation (CSIRO) through funding appropriated by the Australian Government’s Department of Home Affairs. This paper does not reflect any Australian Government policy position. For more information regarding this Project, please refer to \url{https://research.csiro.au/6gsecurity/}.

\bibliography{references}
\bibliographystyle{IEEEtran}

% \newpage

% \section{Biography Section}
% If you have an EPS/PDF photo (graphicx package needed), extra braces are
%  needed around the contents of the optional argument to biography to prevent
%  the LaTeX parser from getting confused when it sees the complicated
%  $\backslash${\tt{includegraphics}} command within an optional argument. (You can create
%  your own custom macro containing the $\backslash${\tt{includegraphics}} command to make things
%  simpler here.)
 
% \vspace{11pt}

% \bf{If you include a photo:}\vspace{-33pt}
% \begin{IEEEbiography}[{\includegraphics[width=1in,height=1.25in,clip,keepaspectratio]{fig1}}]{Michael Shell}
% Use $\backslash${\tt{begin\{IEEEbiography\}}} and then for the 1st argument use $\backslash${\tt{includegraphics}} to declare and link the author photo.
% Use the author name as the 3rd argument followed by the biography text.
% \end{IEEEbiography}

% \vspace{11pt}

% \bf{If you will not include a photo:}\vspace{-33pt}
% \begin{IEEEbiographynophoto}{John Doe}
% Use $\backslash${\tt{begin\{IEEEbiographynophoto\}}} and the author name as the argument followed by the biography text.
% \end{IEEEbiographynophoto}

\vfill

\end{document}